\documentclass[
superscriptaddress,
 amsmath,amssymb,
 aps,
 prx,
twocolumn
]{revtex4-2}

\usepackage{graphicx}
\usepackage{dcolumn}
\usepackage{bm}
\usepackage{braket}
\usepackage{amsthm}
\usepackage{xcolor}

\newcommand{\tens}[1]{^{\otimes #1}}
\newcommand{\phitwo}{\ket{\phi_2}}
\newcommand{\phifour}{\ket{\phi_4}}

\newcommand{\xor}{\oplus}
\newcommand{\Xor}{\bigoplus}
\newcommand{\REP}[1]{\mathrm{REP}(#1)}

\newcommand{\Log}[1]{#1_{\mathrm{L}}}

\usepackage{soul}

\usepackage[hidelinks]{hyperref}
\hypersetup{
  colorlinks   = true,
  urlcolor     = blue,
  linkcolor    = blue,
  citecolor   = blue
}

\theoremstyle{plain}
\newtheorem{thm}{Theorem} 
\newtheorem{cor}{Corollary}

\begin{document}

\title{{Mitigating Temporal Fragility in the XY Surface Code}}

\author{Pei-Kai Tsai}
 \affiliation{Yale University, Department of Applied Physics, New Haven, CT 06511, USA}
 \affiliation{Yale Quantum Institute, Yale University, New Haven, CT 06511, USA}
\author{Yue Wu}
 \affiliation{Yale Quantum Institute, Yale University, New Haven, CT 06511, USA}
  \affiliation{Yale University, Department of Computer Science, New Haven, CT 06511, USA}
\author{Shruti Puri}
 \affiliation{Yale University, Department of Applied Physics, New Haven, CT 06511, USA}
 \affiliation{Yale Quantum Institute, Yale University, New Haven, CT 06511, USA}

\date{\today}

\begin{abstract}
An important outstanding challenge that must be overcome in order to fully utilize the XY surface code for correcting biased Pauli noise is the phenomenon of {\it{fragile temporal boundaries}} that arises during the standard logical state preparation and measurement protocols. To address this challenge we propose a new logical state preparation protocol based on locally entangling qubits into small Greenberger–Horne–Zeilinger-like states prior to making the stabilizer measurements that place them in the XY-code state. We prove that in this new procedure $O(\sqrt{n})$ high-rate errors along a single lattice boundary can cause a logical failure, leading to an almost quadratic reduction in the number of fault-configurations compared to the standard state-preparation approach. Moreover, the code becomes equivalent to a repetition code for high-rate errors, guaranteeing a $50\%$  code-capacity threshold during state preparation for infinitely biased noise. With a simple matching decoder we confirm that our preparation protocol outperforms the standard one in terms of both threshold and logical error rate in the fault-tolerant regime where measurements are unreliable and at experimentally realistic biases. We also discuss how our state-preparation protocol can be inverted for similar fragile-boundary-mitigated logical-state measurement. 
\end{abstract}

\maketitle

\section{Introduction} \label{sec: intro}
Fault-tolerant, scalable quantum computation with noisy physical hardware relies on encoding quantum information in a large number of physical qubits making up an error correcting code. Two important performance metrics for an error correcting code are its threshold, which is the physical error rate below which error correction becomes successful, and the amount of error suppression possible for a given number of qubits. These depend on its {\it distance} which sets the minimum weight of an uncorrectable error, the number of fault-configurations leading to uncorrectable errors, and their likelihood which heavily depends on the details of underlying noise model~\cite{dennis_topological_2002, beverland_role_2019}. 

{\it Biased-Pauli noise} is a common noise model describing many practical qubit architectures in which errors that cause bit-flips are far less likely
than those that only lead to phase-flips~\cite{
shulman_demonstration_2012,
pop_coherent_2014,
mirrahimi_dynamically_2014,
waldherr_quantum_2014,
watson_programmable_2018,
albert_pair_2019,
puri_bias-preserving_2020, 
cong_hardware-efficient_2022}. When only phase-flip noise is present, we say that the noise is infinitely biased. The discovery of native bias-preserving controlled-not gates~\cite{
guillaud_repetition_2019,
puri_bias-preserving_2020,
cong_hardware-efficient_2022,
xu_engineering_2022,
yuan_construction_2022} has driven research towards tailoring codes to be highly effective at correcting biased-Pauli noise~\cite{
aliferis_fault-tolerant_2008,
aliferis_fault_2009,
stephens_high-threshold_2013,
robertson_tailored_2017,
tuckett_ultrahigh_2018,
tuckett_tailoring_2019,
hanggli_enhanced_2020,
huang_fault-tolerant_2020,
tuckett_fault-tolerant_2020,
bonilla_ataides_xzzx_2021,
darmawan_practical_2021,
higgott_subsystem_2021,
guillaud_error_2021,
huang_tailoring_2023,
chamberland_building_2022,
xu_tailored_2023}.
Two leading candidates for such codes are the XY surface code~\cite{tuckett_tailoring_2019}
and XZZX surface code~\cite{bonilla_ataides_xzzx_2021}. These are obtained from the standard CSS surface code by local Clifford-deformation of its stabilizers~\cite{dua_clifford-deformed_2024}. Their favorable properties arise from the underlying symmetry of their stabilizers due to which these codes reduce to repetition codes when noise is infinitely biased (see~\cite{brown_conservation_2023} for further discussion on the role of symmetries in error correction). Thus, these codes have a $50\%$ threshold at infinite bias. Moreover, compared to the planar XZZX surface code, the XY code can also tolerate quadratically higher weight phase errors, making it more desirable for correcting strongly biased noise. 

It is natural to ask if the favorable properties of the XY code persist during logical state preparation, which is the very first step in any quantum algorithm. 
In the standard approach, a logical $X$ state $\ket{\Log{+}}$ is prepared by initializing each physical qubit in the $\ket{+}$ state, followed by a measurement of all the code stabilizers. However, in this process half of the stabilizers cannot be used for detecting phase errors, destroying the symmetry of the XY code, a phenomenon that has been referred to as {\it temporal fragile boundaries}~\cite{higgott_improved_2023}. As a consequence, the code does not reduce to a repetition code under pure phase noise and, as we will show, the threshold at infinite bias degrades to $\sim 11\%$. Moreover, during state preparation the code distance to phase errors scales as $\sqrt{n}$ where $n$ is the number of physical qubits of the code. In contrast, if all the stabilizers could be used for error correction then the code could tolerate more phase-flip errors and its distance to these high-rate errors would be $n$. Thus, fragile boundaries during state preparation cause an overall degradation of the XY code. Similar fragile-boundary-induced degradation also arises during standard approach for logical measurements. 

In this work, we propose a new logical state preparation protocol to mitigate the effect of fragile boundaries. In our protocol, physical qubits are entangled into local two- and four-body Greenberger–Horne–Zeilinger(GHZ) states before being entangled into the surface code state by stabilizer measurements. This initialization pattern allows us to use three-fourths of the stabilizers for correcting phase noise and, as we prove, the code reduces to a single repetition code for phase errors with distance $\sqrt{n}$. Thus a $50\%$ threshold against phase-noise is guaranteed in the state preparation process, a substantial improvement over the standard preparation scheme. Moreover, the fact that there is a single repetition code implies that there are $O(2^{\sqrt{n}})$ fault-configurations. This is almost a quadratic reduction compared to $O(\alpha^{\sqrt{n}})$ fault-configurations in the standard preparation protocol where $3.41\leq \alpha \leq 3.67$~\cite{beverland_role_2019}. Thus, our scheme is able to mitigate the degrading effect of fragile-boundaries.

While our analytical results are limited to the case of pure phase noise, we numerically examine the performance of our scheme under fault-tolerant setting where some bit-flip errors and measurement errors are present. For decoding we use a modified minimum-weight matching (MWPM) algorithm as proposed in~\cite{tuckett_fault-tolerant_2020} and implemented using an open-source library~\cite{wu_qec-playground_2023}. We find that our code is successful at reducing the effects of fragile temporal boundaries at experimentally realistic noise biases. We also discuss how our state-preparation scheme can be inverted for fragile-boundary-mitigated logical state measurement. Finally, we present short-depth circuits for Bell-state preparation that are compatible with the conventional layout of the surface code and that introduce minimum additional noise into the code.

This paper is organized as follows. Section~\ref{sec: XY code} starts with a brief outline of the XY surface code. The standard logical state preparation scheme is described in section~\ref{subsec: standard protocol} and our new scheme along with the main theorems are described in section~\ref{subsec: new protocol}. Finally, section~\ref{sec: results} presents the results of numerical simulations and we conclude with discussion on Bell state preparation and further opportunities in section~\ref{sec: discussion}. 

\section{XY Code} \label{sec: XY code}
\begin{figure}
    \centering
    \includegraphics[width=0.4\textwidth]{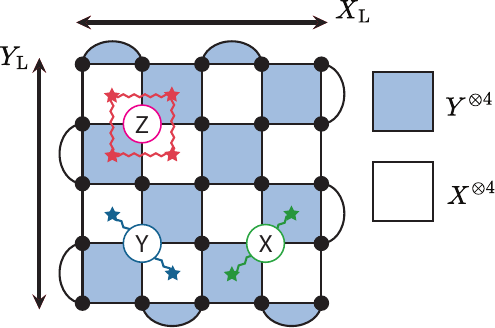}
    \caption{Layout of a rotated XY code, its stabilizers, and error syndromes for high-rate($Z$) and low-rate($X,Y$) errors.}
    \label{fig: XY code}
\end{figure}

We focus on the rotated XY code, also referred to as the tailored surface code~\cite{tuckett_ultrahigh_2018,tuckett_tailoring_2019}, which is defined on a $d \times d$ square lattice for odd $d$ with data qubits on the vertices. The $X$- and $Y$-type stabilizer generators are defined on the faces of the lattice in an alternating checkerboard pattern as shown in Fig.~\ref{fig: XY code}. The $X\;(Y)$-type stabilizers are product of Pauli $X\;(Y)$ operators on data qubits around each face. This $d\times d$ rotated XY code encodes a single logical qubit in $n=d^2$ physical qubits. The $X\; (Y)$-type logical operator is a product of Pauli $X\; (Y)$ acting on qubits on a string connecting the left (top) and right (bottom) boundary. The distance to $X$ and $Y$ errors is $d$. The only non-trivial $Z$-type logical operator is a Pauli $Z$ acting on every qubit~\cite{tuckett_tailoring_2019}. 

$X\; (Y)$ errors anticommute with the $Y\;(X)$-stabilizers creating pairs of syndrome defects oriented diagonally as shown by green (blue) stars in Fig.~\ref{fig: XY code}. A $Z$ error anticommutes with both $X$- and $Y$-type stabilizers leading to four syndrome defects, with pairs on neighboring rows (or columns) oriented vertically (or horizontally), as shown by red stars in Fig.~\ref{fig: XY code}. The underlying structure of syndrome defects generated due to pure $Z$ errors leads to enhanced performance against pure phase noise. More precisely, it has been shown that a $d\times d$ rotated XY code, reduces to a length-$d^2$ or equivalently a length-$n$ repetition code under pure $Z$ noise~\cite{tuckett_tailoring_2019}. Consequently, its threshold to pure $Z$ noise is $50\%$ and the distance to $Z$ errors is exactly the number of qubits $n$. Thus, the code also leads to lower logical error rates under pure $Z$ noise compared to the standard surface code with $O(\sqrt{n})$ $Z$-distance. However, the high $Z$-distance in the XY-code is fragile when $X$ or $Y$ are present~\cite{higgott_improved_2023}. At any one of the four spatial boundaries, $O(\sqrt{n})$ $Z$ errors can combine with a single $X$ or $Y$ error to cause a undetectable logical error. In addition to the spatial fragile boundaries, there are temporal fragile boundaries that occur during state preparation and measurement. While a strategy to mitigate spatial fragile boundary by modifying the stabilizers has been proposed previously~\cite{higgott_improved_2023}, we present the first approach for mitigating temporal fragile boundaries. 

\subsection{Standard Logical State Preparation} \label{subsec: standard protocol}
We first describe the phenomena of temporal fragile boundaries in the standard state preparation approach. For concreteness we consider the preparation of the $\ket{+_\mathrm{L}}$ state but the analysis can be extended to preparation of the $\ket{-_\mathrm{L}},\;\ket{+i_\mathrm{L}},\;\ket{-i_\mathrm{L}}$ states as well. 

The protocol for the preparation of $\ket{+_\mathrm{L}}$ begins by the initialization of each physical qubit in the $\ket +$ state in step 1, followed by the measurement of all the code stabilizers in step 2~\cite{dennis_topological_2002}. The initial unentangled product state $\ket{+}^{\otimes n}$ is the $+1$ eigenstate of $X_\mathrm{L}$ and all the $X$-stabilizers. In the absence of errors, the outcomes of all the $X$-stabilizer measurements are guaranteed to be $+1$, while the $Y$-stabilizer measurements result in outcomes randomly chosen from $(+1,-1)$. Since the stabilizers commute with the logical operators, the qubits after measurements are projected in the $\ket{+_\mathrm{L}}$ state up to a local gauge determined by the outcomes of the $Y$-stabilizer measurements. $Z$ or $Y$ errors can be detected as they anticommute with the $X$-stabilizers and can be corrected. However, because the outcomes of the $Y$-stabilizer measurements are completely random, they cannot be used to detect $Z$ errors. As a result, a $Z$ error only produces two syndrome defects and cannot be differentiated from $Y$ errors. In fact, at this stage the code appears to be identical to the standard surface code with a $\sim 11\%$ threshold and $d=\sqrt{n}$ distance to $Z$-noise on data qubits~\cite{dennis_topological_2002}. Moreover, the number of ways to get a minimum-weight $Z$ error scales as $O(\alpha^{\sqrt{n}})$ where $3.41 \leq \alpha \leq 3.67$~\cite{beverland_role_2019}.

\subsection{New Protocol} \label{subsec: new protocol}
\begin{figure}
    \centering
    \includegraphics[width=0.4\textwidth]{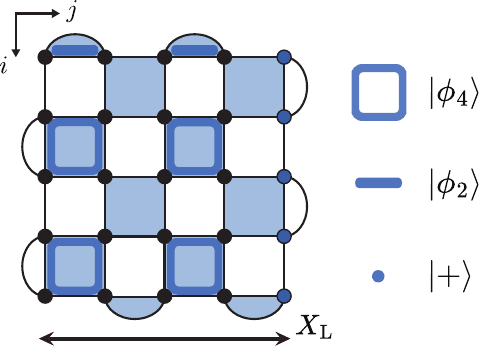}
    \caption{The initialization pattern for $\ket{\Log{+}}$ in our new state preparation protocol. The fixed $Y$ stabilizers are marked by dark blue outline. All the $X$ stabilizers are fixed.}
    \label{fig: initialization pattern}
\end{figure}

We first describe the new protocol to prepare the $\ket{+_\mathrm{L}}$ state which proceeds in two steps. The protocol begins, in step 1, by initializing the qubits in Bell states as indicated in Fig.~\ref{fig: initialization pattern}. The four physical qubits around the $Y$-stabilizer plaquettes, highlighted by dark blue squares, are entangled into the GHZ state $\ket{\phi_4}=\frac{1}{\sqrt2}\left(\ket{+}^{\otimes 4}+\ket{-}^{\otimes 4}\right)$. Each pair of qubits involved in the two-body $Y$-stabilizer along the top $X$-logical boundary, marked by dark blue line, is entangled into the Bell state $\ket{\phi_2}=\frac{1}{\sqrt2}\left(\ket{++} - \ket{--}\right)$ and the remaining qubits along a $Y$-logical boundary, marked as blue dots, are prepared in $\ket{+}$. We will refer to all the qubits except the ones prepared in single-qubit $\ket +$ states as {\it bulk} qubits. Subsequently in step 2 all the stabilizers of the code are measured. 

Note that $\ket{\phi_4}$ and $\ket{\phi_2}$ are respectively the $+1$ eigenstates of $X^{\otimes 4}$ and $X^{\otimes 2}$. Thus the initial state is a $+1$ eigenstate of all the $X$-stabilizers and $X$-type logical operators. Consequently, in the absence of errors, the post-stabilizer-measurement state is the $+1$ eigenstate of the $X$-type logical operators and all the $X$-stabilizer measurement outcomes must be $+1$. 

Importantly, $\ket{\phi_4}$ and $\ket{\phi_2}$ are $+1$ eigenstates of $Y^{\otimes 4}$ and $Y^{\otimes 2}$, respectively. 
Consequently, in the absence of errors, the measurement of outcomes of the marked $Y$-stabilizers must be $+1$. The measurement outcomes of the unmarked $Y$ stabilizers are random $(\pm 1)$. Thus, unlike in the standard protocol, half of the $Y$-stabilizers can be used to detect errors in the new scheme. Note that although only $\ket{\Log{+}}$ is prepared in Theorem~\ref{thm: new}, $\ket{\Log{+i}}$ can be prepared in a similar manner due to the symmetry of the XY code as shown in Appendix~\ref{app: prepare Y_L}. We now state the main theorem.

\begin{thm} \label{thm: new}
{\upshape
In the new state-preparation protocol for the square $d\times d$ $XY$ code, $Z$ errors on all the bulk qubits are correctable (part I) and $Z$ errors on data qubits at the temporal boundary can be decoded as a single repetition code of length $d$ (part II).  }
\end{thm}

\begin{cor}[Fault-configurations] \label{cor: 1}
{\upshape There are $O(2^{\sqrt{n}})$ least-weight fault-configurations due to pure $Z$ errors, where $n=d^2$ is the total number of qubits. This is nearly quadratic improvement over the least-weight fault-configurations in the standard state-preparation approach.}
\end{cor}

\begin{cor}[Threshold] \label{cor: 2}
    {\upshape The threshold to pure $Z$ noise is 50$\%$.}
\end{cor}

\noindent{\bf Proof.}  
Consider the square lattice of the $XY$ code in Fig.~\ref{fig: initialization pattern} where the qubits are placed on the vertices. We will use indices $(i,j)\in \{1,2,...,d\}^2$ to denote the location of data qubits. 
$Z$ errors on the data qubits can be expressed as $Z(\bm z)=\bigotimes_{i,j} (Z_{i,j})^{z_{i,j}}$ with a corresponding binary vector $\bm z=(z_{1,1}, z_{1,2},...,z_{d,d}) \in \{0, 1\}^{d^2}$. $z_{ij}=0\; (1)$ implies no (a $Z$) error on the data qubit located at $(i,j)$. Thus the probability for $z_{i,j}=1$ is equal to the probability of $Z$ errors and the problem of decoding $Z$ errors reduces to correctly determining $\bm z$. 

In the following, we will refer to the $X$ and $Y$ stabilizers with fixed measurement outcome of $+1$ as {\it fixed stabilizers.}
Under pure $Z$ errors, the syndrome measurement of any fixed stabilizer $S$ is $\prod_{(i,j) \in \mathrm{supp}\,S} (-1)^{z_{i,j}}$, where $\mathrm{supp}\,S$ denotes the set of qubits on which $S$ is supported. Thus, it is possible to interpret $\Xor_{(i,j) \in \mathrm{supp}\,S} z_{i,j}=0$ as the parity checks of a classical code where $\Xor$ denotes summation modulo two. A $-1$ outcome of measuring a fixed stabilizer results in the violation of this parity check. The parity checks can be decoded to determine $\bm z$ and the location of $Z$ errors. Next we show that these parity checks reduce to a number of independent classical repetition codes. 

First, consider pairs of qubits on the top row for which we have two-bit parity checks $z_{1,2j-1}\xor z_{1,2j}$, $j=1,2,...,(d-1)/2$ due to the fixed $Y$ stabilizers. Each of the two-bit checks forms a classical 2-bit repetition code $\REP{2}$. By adding the check $z_{1,2j-1}\xor z_{1,2j}$ to the four-bit parity check $z_{1,2j-1}\xor z_{1,2j}\xor z_{2,2j-1}\xor z_{2,2j}$, corresponding to the fixed $X$ stabilizers directly below the top $Y$ stabilizers, we reduce that four-bit parity check to a two-bit parity check $z_{2,2j-1}\xor z_{2,2j}$. Adding this new two-bit parity check to the next four-bit parity arising due to the fixed $Y$ stabilizer in the next row again reduces the latter to a two-bit parity check. In this recursive manner, all four-bit parity checks reduce to two-bit parity checks $z_{i,2j-1}\xor z_{i,2j}$ for $i=1,2,...,d$ and $j=1,2,...,(d-1)/2$ with support on pairs of adjacent qubits in every row.

We can apply this same procedure but this time starting with pairs of qubits on the leftmost column for which we have two-bit parity checks $z_{2i,1}\xor z_{2i+1,1}$, $i=1,2,...,(d-1)/2$ due to the fixed $X$ stabilizers. By adding the check $z_{2i,1}\xor z_{2i+1,1}$ to the four-bit parity check $z_{2i,1}\xor z_{2i,2}\xor z_{2i+1,1}\xor z_{2i+1,2}$, corresponding to the fixed $Y$ stabilizers directly to the right of the $X$ stabilizers, we reduce that four-bit parity check to a two-bit parity check $z_{2i,2}\xor z_{2i+1,2}$. Continuing the recursion, all four-bit parity checks this time reduce to two-bit parity checks $z_{2i,j}\xor z_{2i+1,j}$ with support on pairs of adjacent qubits in every column for $i=1,2,...,(d-1)/2$ and $j=1,2,...,d$. Now consider $z_{2i,2j-1},\; z_{2i,2j},\; z_{2i+1,2j-1},\; z_{2i+1,2j}$, for $i,j=1,2,...,(d-1)/2$, which are supported on qubits around the fixed $Y$ stabilizers. These form a classical 4-bit repetition code $\REP{4}$ with parity checks $z_{2i,2j-1}\xor z_{2i,2j}$, $z_{2i,2j}\xor z_{2i+1,2j}$, $z_{2i+1,2j}\xor z_{2i+1,2j-1}$, and $z_{2i+1,2j-1}\xor z_{2i,2j-1}$. 

Thus we see that for every fixed $Y$ stabilizer there corresponds a classical $\REP{2}$ or $\REP{4}$ code.
A simple counting shows that there are $ (d-1)/2$ $\REP{2}$ and $(d-1)^2/4$ $\REP{4}$ codes. 

At the outset it seems that the probability of successful decoding will be severely limited by the  $\REP{2}$ and $\REP{4}$ codes. However, incorrect decoding of a $\REP{2}$ or $\REP{4}$ results in a flip applied to every bit in its support. Equivalently, this results in $Z$ error applied to every qubit in the support of the corresponding fixed $Y$ stabilizer. However, these qubits are prepared in the entangled states $\ket{\phi_2}$ or $\ket{\phi_4}$ which are eigenstates of $Z^{\otimes 2}$ and $Z^{\otimes 4}$ and are thus invariant under these operators. Hence $Z$ errors on all qubits other than the ones on the last column are correctable. This proves part I of Theorem~\ref{thm: new}.

Finally, we consider the last column of qubits. For this column, each fixed $X$ stabilizer is supported in two qubits. The classical parity checks $z_{i,d}\xor z_{i+1,d}$ ($i=1,2,...,d$) form a classical $d$-bit repetition code $\REP{d}$. Decoding the classical repetition code results in decoding $Z$ errors on these qubits, which is part II of Theorem~\ref{thm: new}. Thus, fewer than $(d-1)/2$-flips on these bits can be corrected. The number of least-weight fault configurations is ${d \choose {(d+1)/2}}=O(2^d)=O(2^{\sqrt{n}})$, which is Corollary~\ref{cor: 1}.  Since the threshold for the classical repetition code is $50\%$, the threshold for $Z$ errors in the quantum code is also $50\%$, from which Corollary~\ref{cor: 2} follows.  

\section{Results} \label{sec: results}
\subsection{Noise model and decoder}
The analysis in the previous section demonstrates the advantage of our scheme over standard state preparation scheme under pure $Z$-noise. In practice there will be some bit-flip noise affecting the qubits. To compare the performance of the two approaches in this experimentally relevant situation we resort to numerical simulation of the state preparation protocol with practical decoding algorithms. We will use a phenomenological model where (a) $X$, $Y$, and $Z$ errors are applied, with probabilities $p_x$, $p_y$, and $p_z$ respectively, on data qubits after they are initialized in the product state $\ket{+}^{\otimes d^2}$ in case of standard scheme or Bell states in case of our proposed scheme, (b) $X$, $Y$, and $Z$ errors are applied, with probabilities $p_x$, $p_y$, and $p_z$ respectively, on data qubits after each round of stabilizer measurements and (c) measurement errors are applied with probability $p_\mathrm{m}$. For fault-tolerance to measurement errors, we perform $d$ rounds of stabilizer measurements after the measurements in step 2. 
Our aim is to estimate the logical error rate as a function of the total probability of errors on the data qubits $p=p_x+p_y+p_z$ for a given bias $\eta=p_z/(p_x+p_y)$. We also assume $p_x=p_y$ for simplicity.

In the standard state-preparation approach, a $Z$ error on the data qubits flips two stabilizers in the measurement round in step 2. In contrast, in our scheme, a $Z$ error on a qubit on the top or right boundary flips two stabilizers while a $Z$ error on any other qubit flips three stabilizers in step 2. A $Y$ error in either schemes flips two stabilizer measurement outcomes. In both the schemes after the first measurement round, a $Z$ error always flips all neighboring stabilizers since all stabilizers can be used for error correction. This implies that standard minimum-weight perfect matching (MWPM) algorithm cannot be used to optimally correct for $Z$-biased noise and instead we use a modified MWPM algorithm introduced in~\cite{tuckett_fault-tolerant_2020}. We refer to this as the Tuckett decoder, which we further adapt to account for the fact that only certain stabilizers can used for error correction in the first measurement round (see Appendix~\ref{app: decoder} for details). 

\begin{figure}
    \centering
    \includegraphics[width=0.4\textwidth]{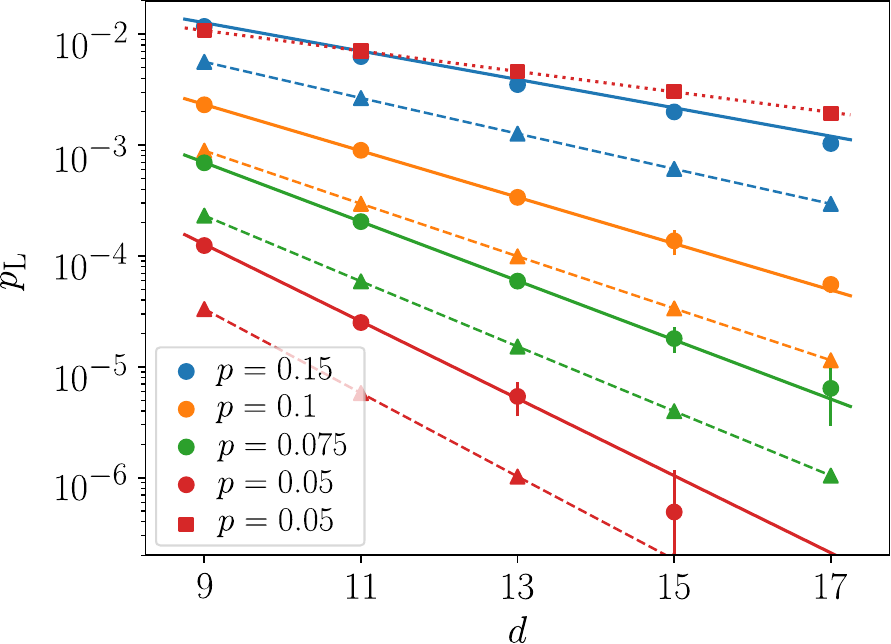}
    \caption{Scaling of logical error rate below threshold for various $p$ with $p_\mathrm{m}=0$. Filled circles show the logical error rates at different $p$ with the new protocol. For comparison, the subthreshold scaling curve for $p=0.05$ with standard protocol is also plotted with square markers. The solid and dotted lines are obtained from linear fits. Filled triangles are the logical error rates of repetition codes with distance $d$ and bit-flip rate $p$. The dashed lines through the triangles are just drawn for easy visualization.}
    \label{fig: subthreshold}
\end{figure}

\subsection{No measurement error, $p_\mathrm{m}=0$}
We first benchmark the adapted Tuckett decoder under ideal measurements $p_\mathrm{m}=0$ and $\eta=\infty$, so that only pure $Z$ noise is present. In this limit we find that the adapted Tuckett decoder results in a $49.1(2)\%$ threshold for our state-preparation scheme and $10.1(1)\%$ threshold for the standard scheme, which are in agreement with our analytical predictions. Next, we analyze the sub-threshold scaling of logical error rate. The filled-circles in Fig.~\ref{fig: subthreshold} show the logical error rate as a function of $d$ for different values of $p$ with $\eta=\infty$ and $p_\mathrm{m}=0$ for the new preparation scheme. For comparison, the filled triangles are the logical error rates for repetition codes of length $d$ as a lower bound of the performance of our protocol. We observe that the numerically obtained logical error rate of our scheme is systematically larger than that for repetition code, indicating that there is scope to further improve over the decoder.

We fit the data to the ansatz $\log \Log{p} = (\alpha d+\beta) \log p + (\gamma d+\delta)$. For our preparation scheme we find the fit parameters to be $(\alpha,\beta,\gamma,\delta)=(0.46(1), 0.04(1), 0.243(9), -0.7(1))$. Recall Theorem~\ref{thm: new} due to which $(d+1)/2$ phase-flip errors occurring on the last column of qubits are uncorrectable so that the logical error rate scales as $p^{(d+1)/2}$ (at low $p$) and $\alpha_\mathrm{ideal}=0.5$. The value of the slope with the adapted Tuckett decoder, $\alpha\sim 0.46$, is close to $\alpha_\mathrm{ideal}$. Despite good agreement with the ideal slope, the adapted Tuckett decoder does not reduce to an ideal decoder as shown with examples in the Appendix~\ref{app: decoder}.

Nonetheless, at fixed $p$, $p_\mathrm{L}$ for our scheme is several orders of magnitude smaller than with the standard preparation scheme, as seen for example, by comparing the logical error rate for $p=0.05$ with the standard preparation scheme (filled red squares) and the new scheme (filled red circles).
For more precise comparison, we examine the error suppression factor $\Lambda$ which determines how fast the logical error rate decreases as the distance increases at a given $p$ ($p_\mathrm{L}=O(\Lambda^{-d})$)~\cite{martinis_qubit_2015}. We obtain $\Lambda$ for $p=0.05$ from the slopes of the fitted straight lines through the red circles and squares respectively for the two protocols. For our protocol we get $\Lambda=2.3(1)$ which is nearly twice of  $\Lambda=1.2(1)$ for the standard protocol, indicating a much faster suppression in logical error rates as larger codes are used with our protocol. The large difference in the values of $\Lambda$ indicates that the number of fault configurations with the adapted Tuckett decoder for our scheme is indeed much smaller compared to that for the standard approach.

Despite the sub-optimality of the Tuckett decoder, it reproduces the analytically predicted thresholds and overall sub-threshold scaling behavior for $\eta=\infty$ fairly well. Thus, we also use it for the finite-$\eta$ case. 
For an experimentally realistic high bias $\eta=10^4$~\cite{xu_engineering_2022,chamberland_building_2022}, the threshold for the total error rate with our scheme is $15.1(2)\%$ in comparison to that of $10.2(1)\%$ with the standard approach. A plot of threshold as a function of the bias for $p_\mathrm{m}=0$ has been shown in Appendix~\ref{app: thresh_no_meas}. 

\begin{figure}
    \centering
    \includegraphics[width=0.4\textwidth]{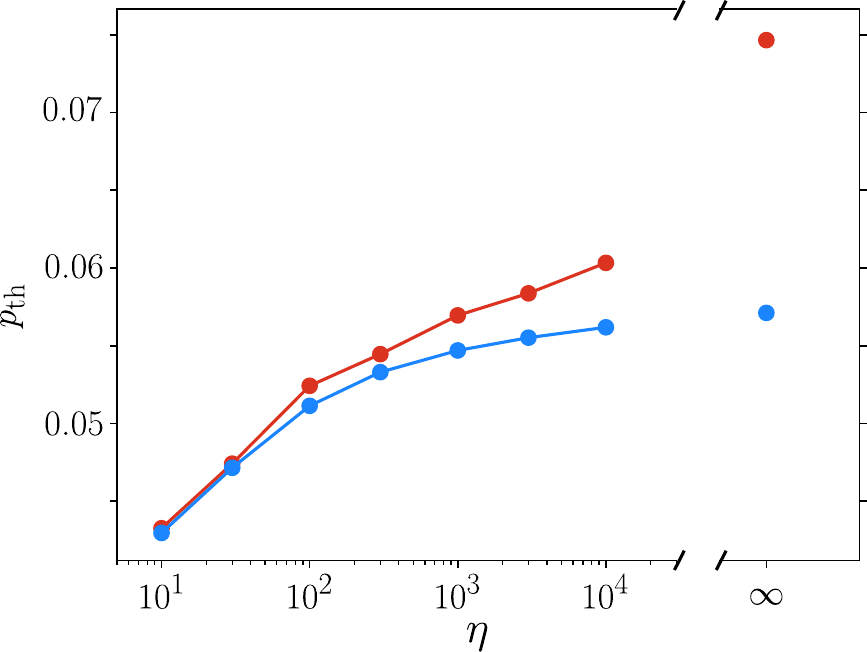}
    \caption{Noise thresholds as a function of bias $\eta$ for the standard state-preparation (blue) and the new protocol (red) for $p_\mathrm{m}=p$.}
    \label{fig: threshold_compare}
\end{figure}

\subsection{With measurement error, $p_\mathrm{m}=p$}
\begin{figure}
    \centering
    \includegraphics[width=0.4\textwidth]{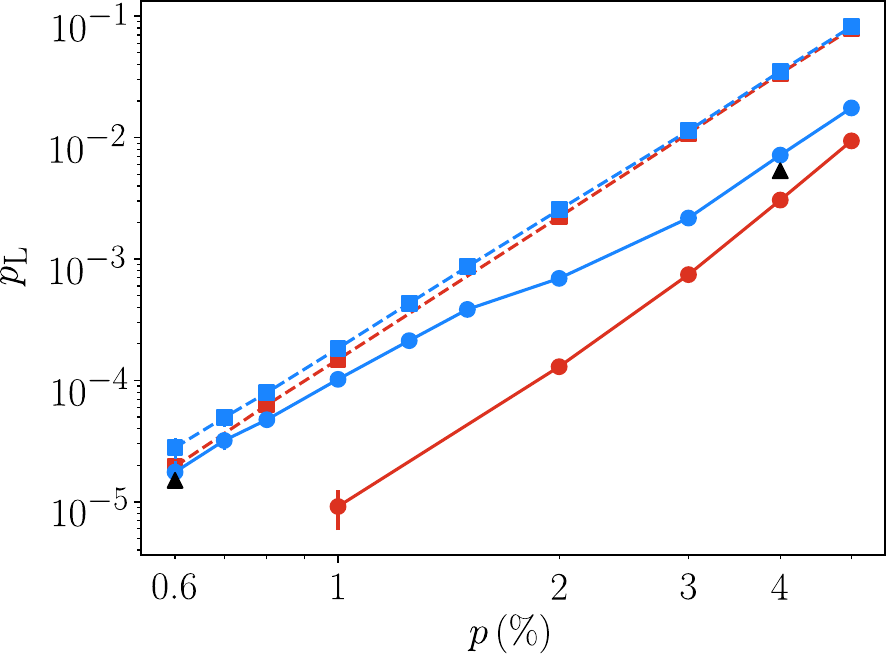}
     \caption{{Logical error rates for the standard (filled-squares) and new preparation (filled-circles) schemes with $p_\mathrm{m}=p$ at $\eta=\infty$ (red) and $\eta=10^4$ (blue) for $d=7$. The solid and dashed lines are shown as guides for the eye and are not obtained from any fits. The black triangles show memory logical error rate at $p=0.6\%, 4\%$ for $\eta = 10^4$.}}
    \label{fig: pL_compare}
\end{figure}

We now consider non-zero measurement errors and assume $p_\mathrm{m}=p$~\cite{tuckett_fault-tolerant_2020}. The plot of threshold as a function of $\eta$ with the two schemes is shown in Fig.~\ref{fig: threshold_compare}. In this case, the threshold difference between the two schemes is less dramatic. At infinite bias the threshold increases from $5.66(1)\%$ for the standard scheme to $7.47(5)\%$ for our scheme, while for $\eta=10^4$ the threshold increases from $5.62(2)\%$ to $6.03(4)\%$. Figure~\ref{fig: pL_compare} compares the logical error rates for the two preparation approaches with $d=7$. For $\eta=\infty$ we find that the logical error rate with our scheme is about an order of magnitude smaller compared to the standard approach. However, the difference between the two approaches becomes smaller as $p$ decreases. 

We attribute this effect to the decreasing contribution from temporal boundaries at smaller $p$ where the gain from our protocol is reduced. For example, at $\eta=10^4$ and $p=4\%$, the state preparation error rate with our approach almost reaches the floor set by the logical memory error rate ($5.3(1)\times 10^{-3}$) for the same parameters shown as black triangle in Fig.~\ref{fig: pL_compare}. The standard preparation scheme clearly cannot reach this floor due to large contribution from temporal boundary errors. On the other hand at $\eta=10^4$ and $p=0.6\%$, the logical memory error rate is $1.5(2)\times 10^{-5}$. Our state preparation reaches this value but the logical error rate with the standard preparation scheme is slightly higher $2.8(2)\times 10^{-5}$. 

These results confirm that our scheme can indeed reduce the amount of additional state-preparation errors due to temporal fragile boundaries. For low physical error rates where the improvement looks less significant, the dominant contribution to logical error rate is mainly from measurement errors and not due to temporal boundaries.

\section{Discussion} \label{sec: discussion}
In this work we proposed a new state-preparation protocol for the XY code which mitigates the effect of fragile temporal boundaries based on using local GHZ states. We also studied the performance of our approach with a practical decoder. In Appendix~\ref{app: logical measurement} we discuss how this protocol can be inverted to realize fragile-boundary-mitigated logical measurements.

\begin{figure}
    \centering
    \includegraphics[width=0.4\textwidth]{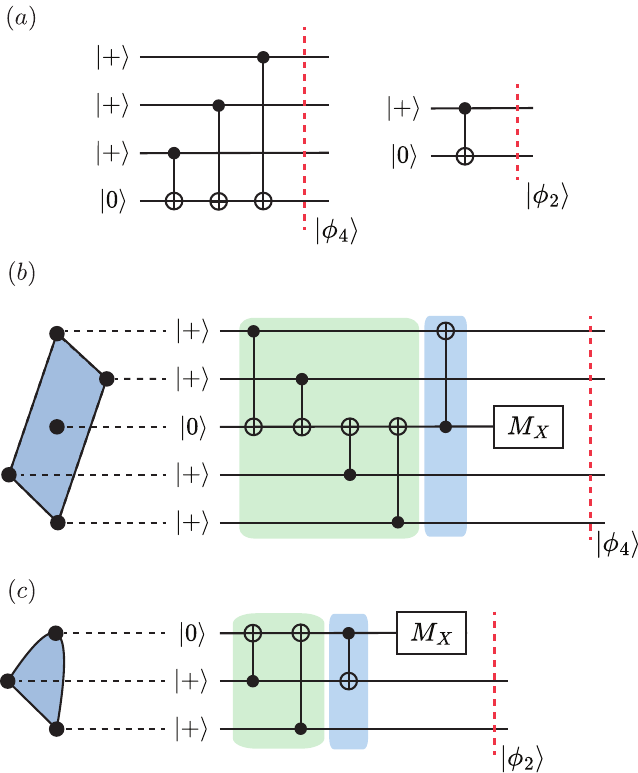}
    \caption{(a) Simple circuits to prepare $\phifour$ and $\phitwo$. (b) An alternative circuit for $\phifour$ which is compatible of standard surface code qubit layout where data qubits only connect to ancilla. (c) A similar circuit to prepare $\phitwo$ on data qubits.}
    \label{fig: XY_code_circuits}
\end{figure}

Practically, it is necessary to be able to prepare the GHZ states with a short-depth circuit in a bias-preserving way. One possible circuit is shown in Fig.~\ref{fig: XY_code_circuits}(a) in which the GHZ states are prepared using CX gates between the data qubits. One drawback of this circuit is that a single high-rate $Z$ error on the data qubit which is the common target for all the CX gates can spread to multiple data qubits causing a correlated error. Moreover, this circuit is also not compatible with the standard connectivity of the surface code where the data qubits don't interact with each other directly, but only interact with an ancilla. 

To overcome these shortcomings, we also give an alternative circuit in Fig.~\ref{fig: XY_code_circuits}(b) and~\ref{fig: XY_code_circuits}(c). This circuit can be effectively understood as first creating a five- (three-) body GHZ state on four(two) data qubits and one ancilla, and then disentangling the ancilla from the remaining data qubits.
Crucially, the ancilla is left in the $|+\rangle$ state at the end in the absence of noise. A single $Z$-type error on the ancilla causes it to end up in the $|-\rangle$ state. Thus a $X$-measurement performed on the ancilla at the very end reveals the presence of $Z$ errors on it. The GHZ state is used in the code only after the ancilla is measured in $\ket{+}$. 
Importantly, this heralding eliminates to first-order error correlations on data qubits caused by $Z$ errors on the ancilla. This implies that the effective noise resulting from this circuit, to first order, can be modeled as independent single-qubit Pauli errors on qubits applied after the circuit which are already accounted for in the phenomenological noise model as independent Pauli noise is applied to qubits after the GHZ state preparation and before the code stabilizers are measured. This ancilla-noise robustness comes at the cost of one extra CX gate compared to the circuit in Fig.~\ref{fig: XY_code_circuits}(a) and compared to the standard stabilizer measurement circuit. Ultimately, future work should consider a full circuit-level simulation of our scheme with additional modifications to mitigate the fragile spatial boundaries~\cite{higgott_improved_2023}.

Moreover, there is considerable room for improving the performance of our scheme by improving the decoder. One possible path would be to combine the Tuckett decoder with belief-propagation~\cite{
criger_multi-path_2018,
acharya_suppressing_2023, 
caune_belief_2023,
higgott_improved_2023}. Ultimately, in order to fully understand the advantages and limits of our scheme a  hypergraph decoder will be necessary. While such a decoder may be inefficient, approximate solutions may be sufficient to reach reasonably low error rates with reasonable latency~\cite{wu_yue_notitle_nodate, wu_hypergraph_2024}.

Other than the XY code, another example where the problem of temporal fragile boundary appears is the tailored XZZX surface code~\cite{xu_tailored_2023} which also reduces to a $n$-bit repetition code at infinite bias as a quantum memory, where $n$ is the number of physical qubits. As $Z_L = Z^{\otimes n}$, the standard way to prepare $\ket{0_L}$ would be to initialize qubits in $\ket{0}^{\otimes n}$. Remarkably however in this case all stabilizer measurement outcomes are random, preventing us from detecting any error occurred during state preparation. Initializing qubits in local entangled states may allow us to fix more stabilizers and thus gain information about errors in the first round. We leave this as an open problem for future work.

\begin{acknowledgements}
This material is based upon work supported by the National Science Foundation (CAREER grant no. 2145223) Any opinions, findings, and conclusions or recommendations expressed in this material are those of the authors and do not necessarily reflect the views of the National Science Foundation. The problem of fragile temporal boundary was also described by Benjamin Brown to SP in personal communication in 2021, although not by that name. We also thank Shilin Huang and Shraddha Singh for useful discussions. 
\end{acknowledgements}

\appendix
\section{State preparation of $\Log{Y}$ eigenstate $\ket{\Log{+i}}$} \label{app: prepare Y_L}
\begin{figure}[ht]
    \centering
    \includegraphics[width=0.4\textwidth]{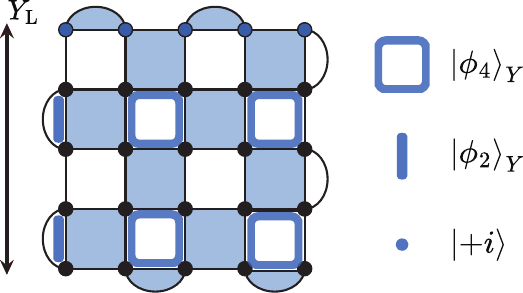}
    \caption{The initialization pattern for $\ket{+i_\mathrm{L}}$ in our new state preparation protocol where $\phifour_Y=\frac{1}{\sqrt 2}\left(\ket{+i}\tens{4} + \ket{-i}\tens{4} \right)$ and $\phitwo_Y=\frac{1}{\sqrt 2}\left(\ket{+i}\tens{2} - \ket{-i}\tens{2} \right)$.}
    \label{fig: XY init Y}
\end{figure}

A similar construction in Fig.~\ref{fig: XY init Y} shows the local initialization pattern which is an eigenstate of $\Log{Y}$. 
Due to the symmetry of $X$ and $Y$ in the XY code, the same argument as in Theorem~\ref{thm: new} applies to prove the $\REP{d}$ structure and 50\% threshold for state preparation. 

\section{Threshold when $p_\mathrm{m}=0$} \label{app: thresh_no_meas}
\begin{figure}[ht]
    \centering
    \includegraphics[width=0.4\textwidth]{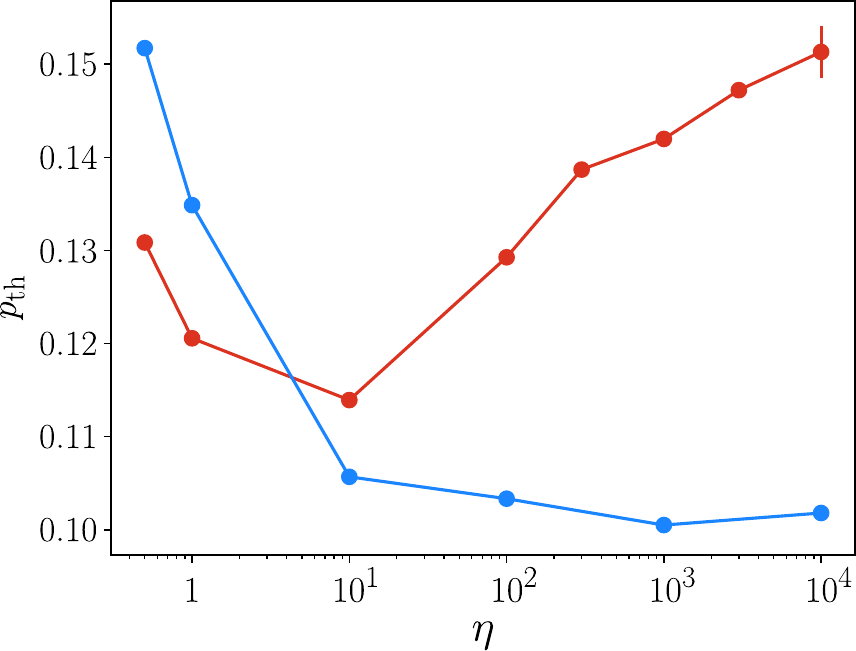}
    \caption{Noise thresholds as a function of bias $\eta$ for the standard state-preparation (blue) and the new protocol (red) for $p_\mathrm{m}=0$.}
    \label{fig: cc_threshold_compare}
\end{figure}

Figure~\ref{fig: cc_threshold_compare} shows threshold as a function of $\eta$ when $p_\mathrm{m}=0$ and Pauli errors applied on data qubits. 

\section{Logical Measurement of $\Log{X}$} \label{app: logical measurement}
The standard protocol for $\Log{X}$ measurement proceeds by measuring each data qubit in the $X$ basis~\cite{dennis_topological_2002}. The measurement outcomes $x_i\in\{0,1\}$ can be added to obtain the logical measurement result $\Log{x} = \Xor_i x_i$. 

In the absence of errors, the measurement outcomes of qubits supported by an $X$-type stabilizer $S$ must sum to zero: $\Xor_{v\in\mathrm{supp}\,S}x_v=0$ since all measurements commute with $S$. The $X$-type stabilizers can thus be effectively used to detect and correct data qubit $Z$ errors (caused by a physical $Z$ or $Y$ error on the data qubit or by a measurement errors). The $Y$-type stabilizers do not provide any information about errors since they don't commute with the data qubit $X$ measurements. Thus, there is no way to detect $X$ errors which is not a problem as these errors don't affect the $X_\mathrm{L}$ measurement anyway. Nonetheless, only half the stabilizers can be used for correcting errors and there is no way to distinguish a $Z$ error from $Y$ error. This results in fragile temporal boundaries similar to the case of state preparation. 

We overcome this challenge by ``inverting" the new state preparation protocol. We measure the local operators that stabilize the Bell states $|\phi_4\rangle$ and $|\phi_2\rangle$ which are  $\{YYYY,XXII,IXXI,IIXX\}$ and $\{YY,XX\}$ respectively. The qubits in the last column are measured in the $X$ basis. 
The result of $\Log{X}$ measurement can be inferred from summing over all disjoint $XX$ and $X$ measurements.
Moreover, the measurement outcomes obey the set of REP(4), REP(2), and REP($d$) parity checks as described under Theorem~\ref{thm: new}. 

We know from the discussion under Theorem~\ref{thm: new} that incorrect decoding of a REP(2) or REP(4) results in $Z\tens 2$ or $Z\tens 4$ applied to qubits supporting $\phitwo$ or $\phifour$. However, $Z\tens 2$ on $\phitwo$ or $Z\tens 4$ on $\phifour$ commutes with operators being measured. Thus we conclude that in this new measurement protocol for the square $d\times d$ $XY$ code, $Z$ errors on data qubits at the temporal boundary can be decoded as a single repetition code REP($d$) on the last column of qubits. It also follows that there are $O(2^{\sqrt{n}})$ least-weight fault-configurations, where $n=d^2$ is the total number of qubits and that the threshold to pure $Z$ noise is 50$\%$.

\section{Adapted Tuckett Decoder} \label{app: decoder}
In this work, we apply the XY code decoder exploiting the symmetries of the code and noise bias~\cite{tuckett_fault-tolerant_2020}. We refer the reader to~\cite{tuckett_fault-tolerant_2020} for details. In this section we only highlight the modifications made to the original decoder for state-preparation. 

The only difference between the original decoder and the decoder we use is how the matching graph is weighed in the first time-step to account for the temporal boundaries. We need to add virtual vertices at the temporal boundaries. 
The vertex for an unfixed stabilizer can be either matched to its virtual temporal vertex with zero weight, or matched to any other vertices with normal weights corresponding to qubit $X$, $Y$, $Z$ errors. The vertex for a fixed stabilizer can be matched to its virtual temporal vertex with weight corresponding to measurement errors $p_\mathrm{m}$.

For the standard preparation approach however, the syndromes due to $Z$ errors are identical to the syndromes due to $Y$ errors in the first time-step. Thus, we modify the Tuckett decoder in this case so that only diagonal edges corresponding to $Y$ errors are allowed in the first time-step with weight corresponding to the probability of $Z$ and $Y$ errors. If we don't do this and instead use parallel edges like in the original Tuckett decoder then the performance of the standard approach degrades substantially.  

\begin{figure}
    \centering
    \includegraphics[width=0.4\textwidth]{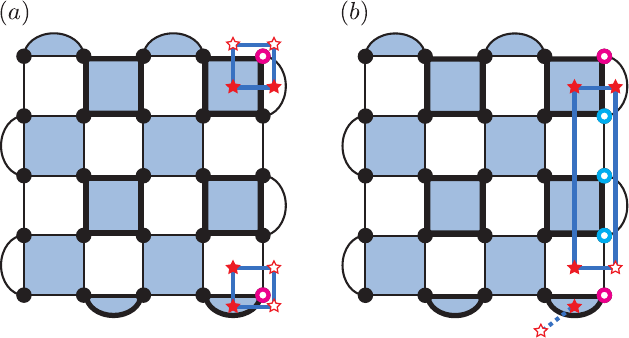}
    \caption{Decoder confusion during state preparation where the decoder fails to correct $(d-1)/2$ phase-flip errors. Note that in (b), the syndrome at the bottom-right corner is connected along time-dimension with zero weight (shown as dashed line). This is because that stabilizer is unreliable at the first stage, so it can be seen as a time-like error.}
    \label{fig: decoder_confusion_state_prep}
\end{figure}

Recall that in the case of state preparation with the new protocol, an optimal decoder should correct up to $(d-1)/2$ $Z$ errors on the last column of qubits, however, with an example shown in Fig.~\ref{fig: decoder_confusion_state_prep} we find that the Tuckett decoder is unable to achieve this. The unfixed stabilizers or stabilizers whose measurement outcome is unknown and cannot be used for error correction are marked with thick black outline in the figure for clarity. Figure~\ref{fig: decoder_confusion_state_prep}(a) shows the  syndromes in filled stars due to two $Z$ errors on qubits marked in red circles. The solid lines shows the possible edges from matching. In this case the decoder assigns $Z$ errors to qubits correctly. However, there is an alternate edge-matching of same weight shown in Fig.~\ref{fig: decoder_confusion_state_prep}(b). In this case the decoder assigns $Z$ errors to qubits marked in solid blue, which differs from the actual $Z$ errors in red by a logical operator. 

\bibliography{references}

\end{document}